\documentclass[prl,twocolumn,floatfix,showpacs,superscriptaddress]{revtex4}

\usepackage{amsmath}
\usepackage{amsfonts}
\usepackage{graphicx}

\DeclareMathSymbol{\N}{\mathalpha}{AMSb}{'116}

\begin{document}

\title{Effective separability of typical entangled many-body states}
\author{S. Camalet}
\affiliation{Laboratoire de Physique Th\'eorique de la Mati\`ere Condens\'ee, UMR 7600, 
Universit\'e Pierre et Marie Curie, Jussieu, Paris-75005, France}
\date{Received: date / Revised version: date }
\begin{abstract}
We consider two systems of harmonically trapped particles in a typical pure state 
of the Hilbert space defined by given values of the particle numbers and energies 
of the two gases. Such a state is entangled but we show that, for large systems, 
the resulting correlations between the two gases are identical to those of a separable 
mixture. This result can be generalized to other physical systems. We discuss 
the relation of this effective separability to the well-known existence of quantum 
correlations in any entangled state. We study in detail a small bipartite system and 
find that its correlations are well explained by the large systems results. 
\end{abstract} 


\maketitle

According to the postulate of equal a priori probabilities, an isolated quantum system 
in equilibrium must be described by a microcanonical mixed state, i.e., a density matrix 
which commutes with the system Hamiltonian and whose only nonzero diagonal 
elements correspond to eigenenergies in a given energy interval. Recently, it has been 
shown that, for many physically relevant properties, a pure state description is also 
possible. An equilibrium state is characterized by its energy and possibly other 
thermodynamic variables such as particle number or volume. Almost all pure states 
in the Hilbert space defined by these parameters lead to expectation values of interest 
identical to those of the corresponding microcanonical mixture, provided the considered 
system is large enough \cite{S,R,PREL}. Most of the effort has been devoted to deriving 
the canonical ensemble for a system S weakly coupled to a large heat bath B from pure 
states of the composite system consisting of S and B \cite{T,GLTZ,GOM,EPJB}. 
But the pure state description of equilibrium is not restricted to the degrees of freedom 
of a subsystem of a larger system. For example, the density profile of a gas in a pure state 
of macroscopically well-defined energy is indistinguishable from that of the corresponding 
microcanonical state \cite{PREL}.  
 
The effective equivalence of pure states with microcanonical states raises an interesting 
question in the context of multipartite systems. Consider two systems $A$ and $B$ in 
a typical state of a Hilbert space characterized by some parameters such as the energies 
of $A$ and $B$. If the above discussed equivalence applies, this entangled state cannot 
be distinguished from the corresponding micronanical state. But this microcanonical 
state is separable, i.e., is a mixture of product states, showing possibly only classical 
correlations between $A$ and $B$. On the other hand, it has been proved that 
the entangled character of any pure state can be revealed by local measurements 
on systems $A$ and $B$, irrespective of their nature and size \cite{GP,PR}. The only 
pure states which do not violate Bell's inequalities \cite{B,CHSH} are product states and 
hence the considered entangled state of $A$ and $B$ must exhibit quantum correlations. 
This seems to be in contradiction with a possible equivalence to a separable state.  

In this Letter, we consider two harmonically trapped gases of bosons or fermions, 
in an entangled state characterized by the energies and particle numbers of the two 
systems. We first show that, for large systems, almost all pure states determined by 
these thermodynamic parameters, lead to the same bipartite correlations. 
These correlations are found to be identical to those of a separable mixture. For small 
particle numbers and energies, the model we study is simple enough to allow 
the complete determination of the basis spanning the corresponding Hilbert space. 
It is thus possible to evaluate correlations between systems $A$ and $B$ for particular 
entangled states drawn from this space. We find that these correlations are well 
described by the expressions obtained for large systems, even for as few as ten 
particles. We discuss in detail the relation of our result to the well-known violation of 
Bell's inequalities mentioned above. 

The system $A$ we consider consists of particles confined in a harmonic 
trap and is described by the Hamiltonian 
\begin{equation}
H_A = \omega_A \sum_{k \ge 0} \left( k+ \frac{1}{2} \right) c^\dag_{Ak} 
c^{\phantom{\dag}}_{Ak}  \label{H}
\end{equation}
where $\omega_A$ is the frequency of the harmonic confining potential and 
$c^\dag_{Ak}$ creates a particle in the single-particle eigenstate $k \in \N$. 
Throughout this paper, we use units in which $\hbar =1 $. The cases of bosons 
and fermions will be treated simultaneously in the following. We assume 
that the system $A$ contains a number of particles $N_A$ and has an 
energy $E_A$. The corresponding eigenstates $| \{ n_{Ak} \} \rangle$ 
of $H_A$ satisfy 
\begin{equation}
\sum_{k \ge 0} n_{Ak} = N_A, \quad 
\sum_{k \ge 0} n_{Ak} k = M_A = \frac{E_A}{\omega_A} - \frac{N_A}{2} ,
\label{NM}
\end{equation}
where $M_A$ is an integer, $n_{Ak} \in \N$ for bosons and $n_{Ak} \in \{0,1\}$ 
for fermions. The Hilbert space spanned by these states is denoted by ${\cal H}_A$ 
and its dimension by $D_A$. The system $B$ is described by a Hamiltonian of 
the form \eqref{H} and is characterized by a particle number $N_B$ and an energy 
$E_B=M_B+N_B/2$. We denote the corresponding eigenstates and Hilbert space by 
$| \{ n_{Bk} \} \rangle$ and ${\cal H}_B$.

\begin{figure}[t]
\centering \includegraphics[width=0.45\textwidth]{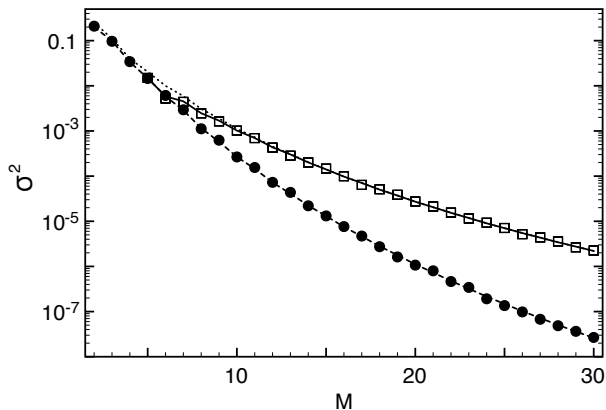}
\caption{\label{fig:DsigmaM} 
Variance $\sigma^2$ as a function of $M$ for bosons, $N_A=N_B=5$, 
$(O_A,O_B)=(O_A^{(0)},O_B^{(0)})$ (solid line and open squares) and 
$(O_A^{(M)},O_B^{(M)})$ (dashed line and filled circles). We remark that $\sigma^2$ vanishes 
for $M \le 4$ in the first case and for $M \le 1$ in the second case. The lines correspond 
to the Hilbert space variance \eqref{sigma}. The symbols are obtained by averaging 
$(\langle O_A O_B \rangle-\langle O_A  \rangle_{E_A} \langle O_B \rangle_{E_B})^2$ over 
$10^3$ typical states \eqref{psi}. For $M>10$, the variance for 
$(O_A,O_B)=(O_A^{(p)},O_B^{(q)})$ where $0 \le p,q \le M$ lies between the two curves 
shown in the figure. The dotted line is $D^{-1}=D_A^{-2}$.
}
\end{figure}

For some practical purposes, it is useful to rewrite the conditions \eqref{NM} as 
\begin{equation}
M_A = \sum_{i=1}^{N_A} k_i \label{NM2}
\end{equation}
where the positive integers $k_i$ obey $k_{i+1} \ge k_i$ for bosons and 
$k_{i+1} > k_i$ for fermions. It is clear from this form that $M_A$ can be as small 
as zero for bosons but $M_A \ge N_A(N_A-1)/2$ for fermions. Another interesting 
conclusion can be drawn from \eqref{NM2} as follows. A fermionic configuration 
$\{ k_i \}$ satisfying \eqref{NM2} with the numbers $N_A$ and $M_A$ corresponds 
to a bosonic configuration $\{ k'_i = k_i - i +1 \}$ satisfying \eqref{NM2} with 
the numbers $N'_A=N_A$ and $M'_A=M_A-N_A(N_A-1)/2$. Consequently, 
the Hilbert space dimension $D_A$ is the same for a fermionic system characterized 
by the numbers $N_A$ and $M_A$ and for a bosonic system characterized by $N_A$ 
and $M'_A$. This dimension increases with $N_A$ and $M_A$. It is equal to $1$ 
for $N_A=1$. For $N_A=2$, it can be easily shown from \eqref{NM2} that, for bosons, 
$D_A=M_A/2 + 1$ for even $M_A$ and $D_A=M_A/2+1/2$ for odd $M_A$. 
The dimension in the fermionic case can be obtained using the fermion-boson 
correspondence just discussed. For larger $N_A$, $D_A$ increases much faster 
with $M_A$, see Fig. \ref{fig:DsigmaM}. The results shown in this figure are obtained 
for $N_A=5$ and are well approximated by $\ln (D_A) \sim \sqrt{M_A}$. For $M_A=30$, 
we find $D_A=674$. A complete discussion of $D_A$ for large $N_A$ and $M_A$ can 
be found in Ref. \cite{GH}.

We consider that the bipartite system consisting of $A$ and $B$ is in a pure 
entangled state
\begin{equation}
| \Psi \rangle = \sum_{\{ n_{Ak} , n_{Bk} \}} 
\Psi_{\{ n_{Ak} , n_{Bk} \}} | \{ n_{Ak} \} \rangle \otimes | \{ n_{Bk} \} \rangle
\label{psi}
\end{equation}
which belongs to the product space ${\cal H}={\cal H}_A \otimes {\cal H}_B$ 
of dimension $D=D_A D_B$. We are interested in the expectation values 
\begin{equation}
\langle O_A O_B \rangle  = \langle \Psi | O_A O_B | \Psi \rangle \label{OAOB}
\end{equation}
where $O_{X}$ ($X=A$ or $B$) is an observable of $X$ with eigenvalues 
between $-1$ and $1$ in a $H_{X}$-invariant space ${\cal H}'_{X}$ 
containing ${\cal H}_{X}$ \cite{fn}. An example is the number of particles of 
system $X$ between two given positions divided by the total number $N_{X}$ 
\cite{PREL}. Bell's inequalities are written in terms of such observables 
\cite{GP,PR,B,CHSH,C}. We now show that, in the limit of large $D$, the expectation 
value \eqref{OAOB} is the same for almost all states $| \Psi \rangle \in {\cal H}$. To do so, 
we use the normalized uniform measure on the unit sphere in ${\cal H}$
\begin{equation}
\mu \left( \left\{ \Psi_{\{ n_\alpha  \}} \right\} \right) = \frac{(D-1)!}{\pi^D}
\delta \Big( 1- \sum_{\{ n_\alpha  \}}  \left| \Psi_{\{ n_\alpha \}} \right|^2 \Big) 
\label{mu}
\end{equation}
where $\{ n_\alpha  \}$ stands for $\{ n_{Ak} , n_{Bk} \}$. Using the expression 
 $\pi^p R^{2p}/ p!$ for the volume of a $2p$-dimensional sphere of radius $R$, 
 we find the Hilbert space average
\begin{equation}
{\overline {\langle O_A O_B \rangle}} = \int d^{2D} \Psi 
\mu \langle O_A O_B \rangle = \langle O_A  \rangle_{E_A} \langle O_B \rangle_{E_B} 
\label{average}
\end{equation}
where $d^{2D} \Psi=\prod_{\{ n_\alpha  \}}  d \mathrm{Re} \Psi_{\{ n_\alpha \}}
d \mathrm{Im} \Psi_{\{ n_\alpha \}}$,  
\begin{equation}
\langle O_{A}  \rangle_{E_{A}} = \frac{1}{D_{A}} 
\sum_{\{ n_{A k} \}} \langle \{ n_{Ak} \}| O_{A} | \{ n_{Ak} \} \rangle \label{micro}
\end{equation}
and $\langle O_{B}  \rangle_{E_{B}}$ is given by a similar expression. 
For the Hilbert space variance 
$\sigma^2={\overline {\langle O_A O_B \rangle^2}}
-{\overline {\langle O_A O_B \rangle}}^2$, 
we obtain, by an analogous calculation,  
\begin{eqnarray}
\sigma^2 &=& \frac{1}{D^2+D} \sum_{\{ n_{\alpha} \}} \sum_{ \{ n'_{\alpha} \}} 
\left| \langle \{ n_{\alpha} \}| O_{A} O_{B}  | \{ n'_{\alpha} \} \rangle \right|^2
\nonumber \\ 
&~&-\frac{1}{D+1} \langle O_A  \rangle_{E_A}^2 \langle O_B  \rangle_{E_B}^2 .
\label{sigma}
\end{eqnarray}
The above sums run only over the configurations satisfying \eqref{NM}. 
An upperbound to the variance $\sigma^2$ is thus obtained by replacing 
one of these sums by a sum over the set of states $|\{ n_{\alpha} \} \rangle$ 
spanning  ${\cal H}'_A \otimes {\cal H}'_B$. Doing so, we find 
$\sigma^2 < D^{-1} \langle O_A^2  \rangle_{E_A} \langle O_B^2 \rangle_{E_B} 
<   D^{-1}$. In conclusion, in the large $D$ limit, $\langle O_A O_B \rangle$ is given 
by the product $\langle O_A  \rangle_{E_A} \langle O_B \rangle_{E_B}$ for almost 
all states $| \Psi \rangle \in {\cal H}$. In other words, for given observables $O_A$ and 
$O_B$, the correlation $\langle O_A O_B \rangle$ for a typical entangled state 
\eqref{psi} is identical to that for the separable state 
$D^{-1} \sum_{\{ n_\alpha \}} | \{ n_\alpha \} \rangle \langle \{ n_\alpha \} |$. 

\begin{figure}[t]
\centering \includegraphics[width=0.45\textwidth]{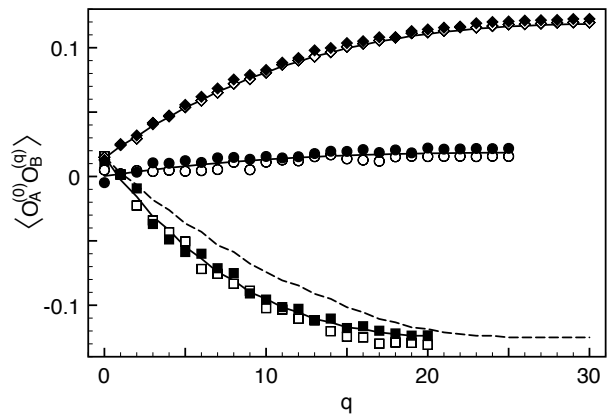}
\caption{\label{fig:AB}  Correlation $\langle O_A^{(0)} O_B^{(q)} \rangle$ as 
a function of $q$ for bosons, $N_A=N_B=5$ and $M=20$ (squares), $25$ (circles) 
and $30$ (diamonds). For each $M$, two typical states \eqref{psi} are shown. 
The lines correspond to the Hilbert space average \eqref{average}. The dashed line 
is this average for fermions, $N_A=N_B=5$ and $M=30$. This curve is flat 
for $q \ge 25$ as $n_k=0$ for $k \ge 25$ in this case. In general, the fermionic curve 
for a given $M$ is close to the bosonic curve for $M'=M-10$ for $0 \le q \le M'$, and 
is flat for $M-5 \le q \le M$. 
} 
\end{figure}

This result seems to be in contradiction with the fact that the entangled nature 
of the state $| \Psi \rangle$ can be revealed by performing local measurements 
on $A$ and $B$. It has been shown that, for any given entangled state of an arbitrary 
bipartite system $(A,B)$, there exist four observables $O_A$, $O'_A$, $O_B$ and $O'_B$ 
with eigenvalues $\pm 1$ such that $|\langle F \rangle| > 2$ where 
$F = O_A (O_B + O'_B) + O'_A (O_B - O'_B)$ \cite{GP,PR}. On the other hand,  the above 
argumentation can be applied to $F$ and one finds, in the large $D$ limit, 
$\langle F \rangle=\langle O_A  \rangle_{E_A} \langle O_B + O'_B \rangle_{E_B}+
\langle O'_A  \rangle_{E_A} \langle O_B - O'_B \rangle_{E_B}$   which is clearly 
between $-2$ and $2$. In short, for any given observables $O_A$, $O'_A$, $O_B$ and 
$O'_B$, $|\langle F \rangle| \le 2$ for almost all entangled states \eqref{psi} but, for each 
of these states, there exist purpose-built observables for which this inequality is violated. 
For large many-body systems, these very special observables might be difficult to measure 
as the number of measurements that can be made in practice is limited. For example, 
measurements of single-particle observables 
$\sum_{k,k'} \lambda_{kk'} c^\dag_{X k} c^{\phantom{\dag}}_{X k'}$ such as particle number 
densities, can be achieved, whereas $n$-particle observables with $n$ of the order of 
$N_{X}$, are practically inaccessible. Another point is worth mentioning here. In the following, 
we consider an observable $F$ such that ${\overline {\langle F \rangle}} \rightarrow 2^-$ 
in the large $D$ limit. In such a case, although 
${\overline {\langle F \rangle^2}}-{\overline {\langle F \rangle}}^2 \rightarrow 0$, the proportion 
of states \eqref{psi} such that $\langle F \rangle > 2$, does not necessarily vanish in this limit 
and the above conclusion may be not strictly correct. However, the proportion of 
states $|\Psi \rangle$ such that $\langle F \rangle > 2+\epsilon$ where $\epsilon$ is a positive 
number, as small as we please, obviously vanishes.  

The effective equivalence of a typical state \eqref{psi} with a separable state can be 
generalized to other systems. First of all, for two systems $A$ and $B$ described by 
the Hamiltonian \eqref{H}, other Hilbert spaces ${\cal H}_{X}$ can be considered. 
The above derivation remains valid, for example, if the condition \eqref{NM2} is replaced 
by $\sum_{i=1}^{N_A} k_i < M_A$ or, in other words, for states $|\Psi \rangle$ given 
by \eqref{psi} with the sum running over all the eigenstates $| \{ n_{Ak} \} \rangle$ and 
$| \{ n_{Bk} \} \rangle$ corresponding to eigenenergies lower than $E_A$ and $E_B$, 
respectively. For other systems, such states show effective separability in a proper 
thermodynamic limit. It can be seen as follows. In this limit, the average \eqref{micro} 
and $S_A=k_B \ln(D_A)$ where $k_B$ is the Boltzmann constant, are, respectively, 
the usual microcanonical average and entropy \cite{Diu}. As is well known, the entropy 
$S_A$ is extensiveand hence the variance \eqref{sigma} vanishes exponentially with 
the system size. 

\begin{figure}[t]
\centering \includegraphics[width=0.45\textwidth]{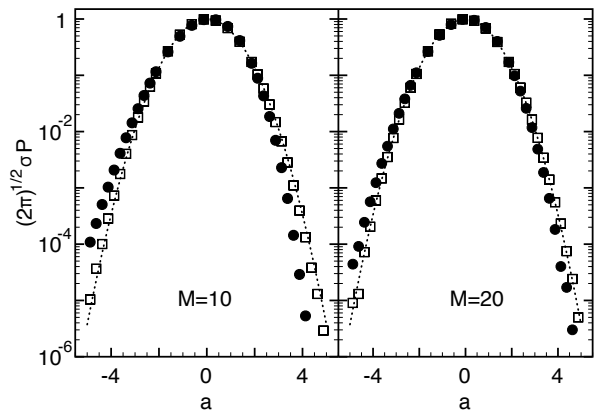}
\caption{\label{fig:dis} Distribution $P\left( \langle O_A O_B \rangle \right)$ as a 
function of 
$a=(\langle O_A O_B \rangle-\langle O_A  \rangle_{E_A} \langle O_B \rangle_{E_B})/\sigma$ 
for bosons, $N_A=N_B=5$, $(O_A,O_B)=(O_A^{(0)},O_B^{(0)})$ (open squares) and 
$(O_A^{(M)},O_B^{(M)})$ (filled circles), and $M=10$ and $20$. The dotted lines 
correspond to the normal distribution of variance $\sigma^2$.  
} 
\end{figure}

We now study in detail a small bipartite system. We consider two systems described 
by the Hamiltonian \eqref{H} with $N_A=N_B=5$ and $M_A=M_B=M$ between 
$0$ and $30$ for bosons and between $N_A(N_A-1)/2=10$ and $40$ for fermions. 
The corresponding eigenstates are determined using \eqref{NM2}. As observables 
$O_X$, we use the operators defined by 
$O_X^{(q)}| \{ n_{Xk} \} \rangle= \pm | \{ n_{Xk} \} \rangle$ with the upper sign 
if $n_{Xq}=0$ where $q$ is a given positive integer, and the lower if  $n_{Xq} \ne 0$. 
For these observables, the first term of \eqref{sigma} simplifies to $1/(D+1)$. 
We remark that $O_X^{(q)}=1-2c^{\dag}_{Xq}c^{\phantom{\dag}}_{Xq}$ for fermions. 
To draw a normalised state \eqref{psi} from the uniform distribution \eqref{mu}, 
we generate $D$ random complex numbers $\Phi_{\{ n_\alpha \}}$ with standard 
normal distribution and then compute the components 
$\Psi_{\{ n_\alpha \}}=\Phi_{\{ n_\alpha \}}/\sum_{\{ n_\alpha \}} |\Phi_{\{ n_\alpha \}}|^2$ 
\cite{GLTZ}. For each state, we evaluate correlations \eqref{OAOB}. Results obtained in 
this way are shown in Fig. \ref{fig:AB}. The agreement with the microcanonical average 
\eqref{average} is excellent. As discussed above, the dispersion of  $\langle O_A O_B \rangle$ 
around this mean value decreases with increasing $M$, see Fig. \ref{fig:DsigmaM}. 
Distributions of correlations $\langle O_A^{(q)} O_B^{(q')} \rangle$ are shown in 
Fig. \ref{fig:dis}. They are constructed by evaluating such expectation values for $10^7$ 
typical states $| \Psi \rangle$. It can be shown that they converge to normal distributions 
as $M$ is increased. 

\begin{figure}[t]
\centering \includegraphics[width=0.45\textwidth]{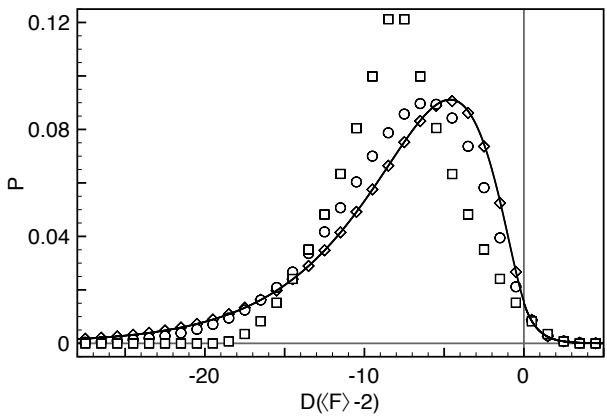}
\caption{\label{fig:F}  Distribution $P(\langle F \rangle)$ as a function of 
$D(\langle F \rangle-2)$ for bosons, $N_A=N_B=5$ and $M=2$ (squares), $3$ (circles) 
and $10$ (diamonds). The line is the large $D$ expression \eqref{P}. 
}
\end{figure}

To discuss Bell's inequalities, let us consider two particular states $|\{n_{Ak}^\pm \}\rangle$ 
satisfying \eqref{NM}, and define observables $\sigma^z_A$ and $\sigma^x_A$ which act 
on the states $|\{n_{Ak} \}\rangle \ne |\{n_{Ak}^\pm \}\rangle$ like the identity operator, and 
are represented by the Pauli matrices $\sigma_z$ and $\sigma_x$, respectively, in 
the basis $\{ |\{n_{Ak}^+ \}\rangle, |\{n_{Ak}^- \}\rangle \}$ of the corresponding subspace. 
For system $B$, we define the observables $\sigma^u_B$ and $\sigma^v_B$ which coincide 
with the unit operator except on the subspace spanned by two states $|\{n_{Bk}^\pm \}\rangle$ 
in which $\sigma^u_B=2^{-1/2}(\sigma^z_B - \sigma^x_B)$ and 
$\sigma^v_B=2^{-1/2}(\sigma^z_B + \sigma^x_B)$ where $\sigma^z_B$ and $\sigma^x_B$ 
are the analogues of $\sigma^z_A$ and $\sigma^x_A$, respectively. We are interested in 
the expectation value
\begin{equation}
\langle F \rangle = \left\langle \sigma^x_A (\sigma^u_B+\sigma^v_B) + 
\sigma^z_A (\sigma^u_B-\sigma^v_B) \right\rangle 
\end{equation}
in a typical state \eqref{psi}. It can be written in terms of the four components 
$\Psi_{\{ n_\alpha \}}$ corresponding to $|\{n_{Ak}^\pm \} , \{n_{Bk}^\pm \}\rangle$. 
To simplify the following expressions, we denote these components by $\Psi_{\pm\pm}$. 
We first observe that, for states $| \Psi \rangle$ such that $\Psi_{++}=\Psi_{--}=0$ and 
$\Psi_{+-}=\Psi_{-+}$, $\langle F \rangle=2\sqrt{2}\eta+2(1-\eta)$ where 
$\eta=2|\Psi_{+-}|^2 \in [0,1]$, and hence varies between $2$ and $2\sqrt{2}$ which is 
the maximum possible value for such an expectation value \cite{C}. So, the CHSH inequality 
\cite{CHSH}, $|\langle F \rangle| \le 2$, is violated by some states \eqref{psi}. For the Hilbert 
space average, we find ${\overline {\langle F \rangle}} = 2-8/D <2$. It is thus interesting 
to study the distribution $P(\langle F \rangle)$ resulting from the measure \eqref{mu}. Figure 
\ref{fig:F} shows such distributions constructed by evaluating $\langle F \rangle$ for $10^8$ 
states $|\Psi \rangle$. We remark that $P$ is independent of the choice of the states 
$|\{n_{Ak}^\pm \}\rangle$, it depends only on the dimension $D$. A large $D$ expression 
for $P$ can be derived as follows. For large $D$, the distribution of the four components 
$\Psi_{\pm\pm}$ is essentially Gaussian. Consequently, using 
$\delta(x)=\int dk \exp(ikx)/2\pi$, $P$ can be written as the Fourier transform of a Gaussian 
integral which is readily evaluated. Then a residue calculation gives 
\begin{eqnarray}
P &\simeq& \frac{D}{8} \frac{e^{-x(\sqrt{2}+1)}}{3\sqrt{2}+4} \Theta (x) \label{P} \\
&~&+ \frac{D}{8} \left( \frac{e^{x(\sqrt{2}-1)}}{3\sqrt{2}-4} + 2 e^{x} (x -2) \right) \Theta (-x) 
\nonumber
\end{eqnarray}   
where $x=D(\langle F \rangle-2)/2$. This expression agrees very well with the results 
obtained for $M$ as small as $10$, see Fig.\ref{fig:F}. From it, we infer that the proportion 
of states \eqref{psi} such that $\langle F \rangle > 2$, is $(40+28\sqrt{2})^{-1} \simeq 0,013$.

In summary, we have studied two harmonically trapped gases in an entangled pure state 
characterized by the particle numbers and energies of the two systems. For almost all such 
states, the correlations between the two systems are identical to those of a separable mixed 
state. We have proved this for large systems by evaluating the Hilbert space average and 
variance of such a correlation, and have shown how this proof can be applied to other 
physical systems. We have also studied in detail a small bipartite system and found that 
its correlations are well explained by the large systems results. To discuss the seeming 
inconsistency between the effective separability found here and the unavoidable violation 
of Bell's inequalities, we have considered an observable which leads to a maximal violation 
of the CHSH inequality \cite{CHSH} for some of the considered states. The proportion of states 
which do not satisfy this inequality remains finite in the limit of large systems but the inequality 
violation is less and less pronounced as the system size is increased.


\begin{thebibliography}{99}

\bibitem{S} A. Sugita, Nonlin. Phenom. Compl. Syst. {\bf 10}, 192 (2007).

\bibitem{R} P. Reimann, Phys. Rev. Lett. {\bf 99}, 160404 (2007).

\bibitem{PREL} S. Camalet, Phys. Rev. E {\bf 78}, 061112 (2008); 
Phys. Rev. Lett. {\bf 100}, 180401 (2008).

\bibitem{T} H. Tasaki, Phys. Rev. Lett. {\bf 80}, 1373 (1998). 

\bibitem{GOM} J. Gemmer and G. Mahler, Eur. Phys. J. B {\bf 31}, 249 (2003); 
J. Gemmer, A. Otte and G. Mahler, Phys. Rev. Lett. {\bf 86}, 1927 (2001).

\bibitem{GLTZ} S. Goldstein, J.L. Lebowitz, R. Tumulka and N. Zangh\`i, 
Phys. Rev. Lett. {\bf 96}, 050403 (2006). 

\bibitem{EPJB} S. Camalet, Eur. Phys. J. B {\bf 61}, 193 (2008).

\bibitem{GP} N. Gisin and A. Peres, Phys. Lett. A {\bf 162}, 15 (1992).

\bibitem{PR} S. Popescu and D. Rohrlich, Phys. Lett. A {\bf 166}, 293 (1992).

\bibitem{B} J.S. Bell, Physics {\bf 1}, 195 (1964).

\bibitem{CHSH} J.F. Clauser, M. A. Horne, A. Shimony and R.A. Holt, 
Phys. Rev. Lett. {\bf 23}, 880 (1969).

\bibitem{GH} S. Grossmann and M. Holthaus, Phys. Rev. E {\bf 54}, 3495 (1996); 
Phys. Rev. Lett. {\bf 79}, 3557 (1997). 

\bibitem{fn} ${\cal H}'_{X}$ is also invariant under the action of $O_X$. 

\bibitem{C} B.S. Cirel'son, Lett. Math. Phys. {\bf 4}, 93 (1980).

\bibitem{Diu} B. Diu, C. Guthmann, D. Lederer and B. Roulet, 
{\it Physique statistique} (Hermann, Paris, 1989).

\end{thebibliography}
\end{document}